\documentclass[10pt,twocolumn,letterpaper]{article}

\usepackage{cvpr} 
\usepackage{times}
\usepackage{epsfig}
\usepackage{graphicx}
\usepackage{amsmath}
\usepackage{amssymb}
\usepackage{booktabs} 
\usepackage{array}  
\usepackage{xcolor,colortbl}
\usepackage{multirow}

\begin{document}

\title{SecureSpeech: \\ Prompt-based Speaker and Content Protection}

\author{Belinda Soh Hui Hui\\
Sinagpore Institute of Technology, Singapore\\
{\tt\small 1900900@sit.singaporetech.edu.sg}
\and
Xiaoxiao Miao\thanks{Xiaoxiao Miao is the corresponding author and this work was conducted while she was at SIT. \\ This study is supported by JST, PRESTO Grant JPMJPR23P9, Japan and Ministry of Education, Singapore, under its Academic Research Tier 1 (R-R13-A405-0005).  }\\
Duke Kunshan University, China\\
{\tt\small xiaoxiao.miao@dukekunshan.edu.cn}
\and
Xin Wang\\
National Institute of Informatics, Japan\\
{\tt\small wangxin@nii.ac.jp}
}

\maketitle
\thispagestyle{empty}

\begin{abstract}
Given the increasing privacy concerns from identity theft and the re-identification of speakers through content in the speech field, this paper proposes a prompt-based speech generation pipeline that ensures dual anonymization of both speaker identity and spoken content. This is addressed through
1) generating a speaker identity unlinkable to the source speaker, controlled by descriptors, and 
2) replacing sensitive content within the original text using a name entity recognition model and a large language model.
The pipeline utilizes the anonymized speaker identity and text to generate high-fidelity, privacy-friendly speech via a text-to-speech synthesis model.
Experimental results demonstrate an achievement of significant privacy protection while maintaining a decent level of content retention and audio quality.
This paper also investigates the impact of varying speaker descriptions on the utility and privacy of generated speech to determine potential biases.
\end{abstract}

\section{Introduction}
\label{sec:intro}

\begin{figure*}[t]
    \centering
    \includegraphics[width=0.9\textwidth]{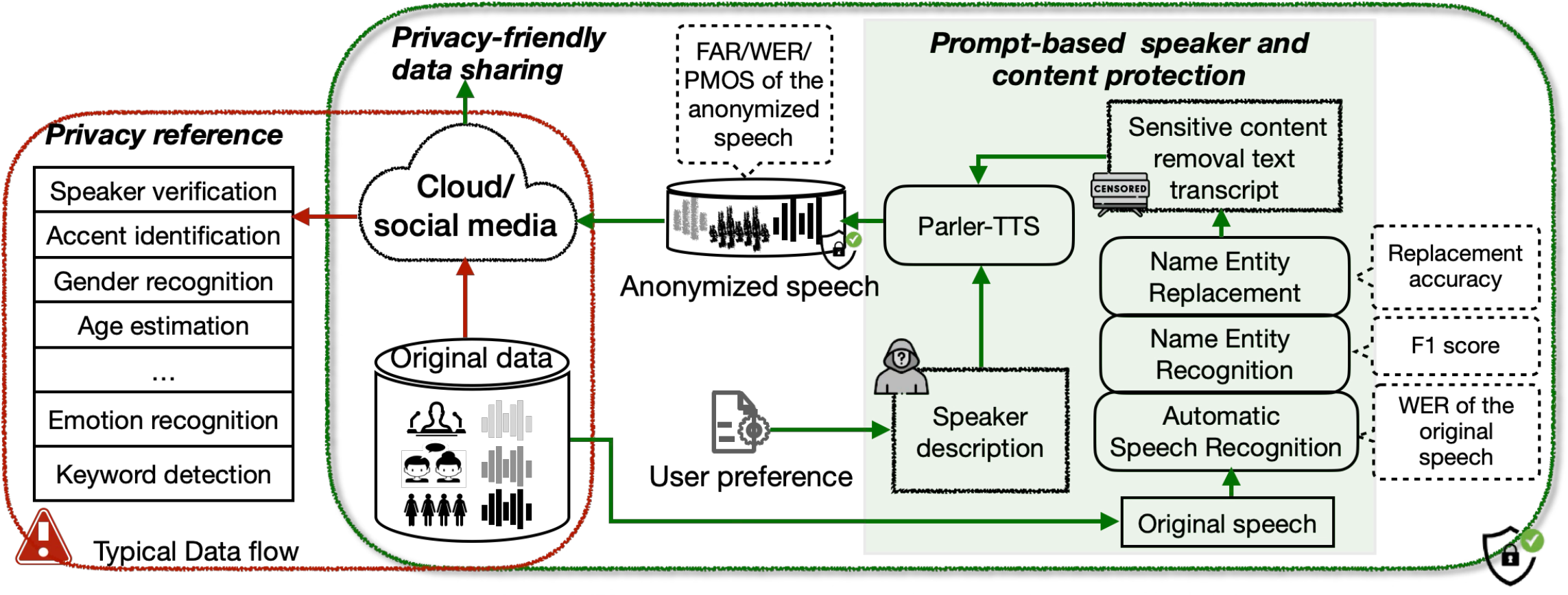}
    \caption{Speech data flow comparison: a typical method (left) and the proposed prompt-based speaker and content anonymization method (right).}
    \label{fig:Illustration of Pipeline}
    \vspace{-3mm}
\end{figure*}

In traditional voice interaction systems, audio data is typically uploaded to a cloud service or shared on social media for further analysis as shown on the left side of Figure \ref{fig:Illustration of Pipeline}. However, speech data has rich information beyond the linguistic content,  encompassing paralinguistic attributes like age, emotion, identity, geographical origin, and health status. Failure to implement voice privacy protection and directly sharing raw speech data on social platforms or with third-party companies may result in privacy leakage \cite{tomashenko2024voiceprivacy}.

One significant privacy concern is the potential for identity theft. State-of-the-art (SOTA) speech generation systems can clone a person's voice using an original speech sample as short as a few seconds \cite{arik2018neural,zhang2023speak} and cloned audio can be challenging for both humans and machines to distinguish from real audio. A survey revealed that participants mistook AI-generated voices as real 80\% of the time and identified them correctly with 60\% accuracy \cite{barrington2024people}. An existing solution in the academic community is user-centric speaker anonymization \cite{tomashenko2021voiceprivacy, vpc2022sum, tomashenko2024voiceprivacy}. This approach processes raw audio data before data sharing to produce anonymized speech, aiming to conceal a speaker’s identity while preserving the original content and naturalness. 

Another significant privacy concern is the potential re-identification of speakers through content information, as speech content may include words that reveal a speaker's identity, such as phone numbers and addresses. Even when a speaker's name is not explicitly labeled, there remains a risk of linking specific audio samples to the speaker.
To address this, researchers have proposed speech content anonymization methods designed to remove sensitive information from audio data \cite{williams2021revisiting, veerappan2024speedf, justin2015speaker}. However, most of these methods explore simplistic techniques, such as bleeping or segment removal, which degrade the utility of the anonymized speech.

Several studies combine speaker and content anonymization to enhance voice privacy. One such example is Speech Sanitizer \cite{qian2019speech}, which performs content anonymization through sensitive keyword detection, spotting, and substitution, and protects speakers using digital signal processing-based voice conversion with optimized warping factors.
Another similar work is SpeechHide \cite{hu2022speechhide}, which anonymizes sensitive content using Named Entity Recognition (NER) and keyword extraction while anonymizing the voiceprint using differential privacy. The limitations of these works include reduced speech intelligibility and naturalness in anonymized speech, as well as increased complexity.
Another simpler pipeline involves transcribing speech into text using an automatic speech recognition (ASR) model, followed by named entity recognition and replacement, and then using a text-to-speech (TTS) system that re-synthesizes speech from the sensitive-removed transcription \cite{xinyuan24_spsc}. While some researchers have previously argued that this ASR + NER + TTS approach perfectly conceals the speaker's identity, it compromises the utility of the speech, such as the unavoidable ASR encoding error, NER detecting error, and the limited variability of synthesized speech produced by the TTS system, especially when the system can generate only a few voices \cite{shamsabadi2022differentially}. However, rapid advancements in ASR, NER, and TTS techniques in recent years have opened new possibilities.

This paper proposes a novel ASR + NER + TTS pipeline for dual anonymization of speaker identity and content\footnote{Code sample is available at https://github.com/pixelable/SecureSpeech.git}\footnote{Audio sample is available at  https://pixelable.github.io/SecureSpeech/}. It leverages a large-scale language model (LLM), Llama-3.2 \cite{petroni2019language}, and a prompt-based TTS system, Parler-TTS \cite{lacombe-etal-2024-parler-tts, lyth2024natural}, as shown on the right side of Figure \ref{fig:Illustration of Pipeline}, allowing users to customize anonymized speaker descriptions and content replacement formats for diverse use cases while ensuring privacy, intelligibility, and naturalness. As of the time of writing, this work is the first to explore prompt-based edits of speaker attributes for both speaker and content anonymization. Meanwhile, we conduct an in-depth analysis of how different prompt-based attributes affect the anonymization performance.

Specifically, for speaker anonymization, most systems replace or transform speaker embeddings with anonymized ones. These anonymized speaker embeddings are typically generated either based on a random selection of dissimilar speaker embeddings~\cite{vpc2022sum} or produced via a blackbox deep neural network~\cite{miao2023language}, limiting users' ability to intuitively control how the anonymized voice sounds. In contrast, our proposed approach utilizes text-based speaker descriptions to directly control the pseudo-speaker attributes in the anonymized speech. 
For content anonymization, the system combines a NER model with Llama-3.2 model to identify and replace sensitive content\footnote{This paper assumes that `sensitive content' corresponds to recognized named entities in the text.}. Then, Parler-TTS takes the speaker description and text transcript with sensitive words removed as input to generate privacy-friendly speech, effectively safeguarding user privacy while ensuring high-quality speech synthesis. 

Additionally, we investigate how speaker descriptions impact the privacy and utility of generated speech, exploring whether biases emerge by adjusting speaker descriptions in terms of gender, pitch, accent, speaking rate, or channel condition. These insights help guide users in crafting effective speaker description prompts, ensuring the speech preserves privacy, maintains high quality, and remains suitable for downstream applications.

\section{Prompt-based Speaker and Content  Anonymization Pipeline}
In this section, we describe the proposed anonymization pipeline, as illustrated on the right side of Figure \ref{fig:Illustration of Pipeline}. The system enables speech-to-speech generation with dual-layer anonymization to protect both speaker identity and spoken content. Key components include speech transcription,  name entity recognition and replacement, speaker description generation, and speech synthesis. Each of these components will be described in the following sections.

\subsection{Spoken Content Protection}
To safeguard spoken content, the system begins with ASR, which transcribes the input speech into text. 
The transcribed text is then processed by a NER model to identify sensitive information such as personal names, locations, and other private entities. Once these sensitive entities are detected, an open-weight LLM is prompted to replace the identified entities with generic or contextually appropriate alternatives. This ensures that the content of the speech is sanitized while maintaining coherence and contextual relevance.
As these models represent well-established techniques, we leverage knowledge from existing SOTA models. 
Note that these are off-the-shelf models, offering users flexibility in replacement. 

\subsubsection{ASR Model - wav2vec2.0 + CTC}
\label{sec:asr}
The ASR model\footnote{\url{https://huggingface.co/speechbrain/asr-wav2vec2-librispeech}} we used is fine-tuned on \textit{LibriSpeech-train-960} with the \textit{wav2vec2-large-960h-lv60-self} model\footnote{\url{https://huggingface.co/facebook/wav2vec2-large-960h-lv60-self}} using a \textit{SpeechBrain} \cite{speechbrain} recipe. Specifically, the \textit{wav2vec2-large-960h-lv60-self} model \cite{wav2vec2} processes raw audio input into latent speech representations using a multi-layer convolutional encoder, followed by a Transformer that converts the latent speech representations into contextual embeddings. This pre-trained model is further fine-tuned for speech recognition using a character-level Connectionist Temporal Classification (CTC) objective \cite{hannun2017sequence}, with two linear layers added to map the contextual embeddings to their corresponding transcriptions.

\subsubsection{NER Model - DeBERTa-L}
\label{sec:ner}
We follow the approach described in \cite{shon2022slue} to implement an NER pipeline. The transcription produced by the ASR system (described in Section \ref{sec:asr}) is processed by a pre-trained DeBERTa-L model, which has 24 transformer layers and a linear classification layer on top for NER. The model is fine-tuned for token classification to detect named entities, their corresponding types, and timestamps within a given sentence.

\subsubsection{Name Entity Replacements Model - Llama-3.2}
\label{sec:ner-replace}

There are several ways to perform named entity replacement after obtaining named entity tags and timestamps. Current content protection studies typically erase named entities or replace them using a predefined replacement list \cite{hu2022speechhide, qian2019speech}. In our approach, we effectively leverage Llama-3.2\footnote{\url{https://huggingface.co/meta-llama/Llama-3.2-1B}}, an open-source LLM, through prompt engineering. By modifying the prompts, we can easily replace named entities with various alternatives, such as dissimilar words, or use a fixed mapping list to guide the language model in generating specific replacements.

\begin{figure}[t]
    \centering
   \includegraphics[width=1\linewidth]{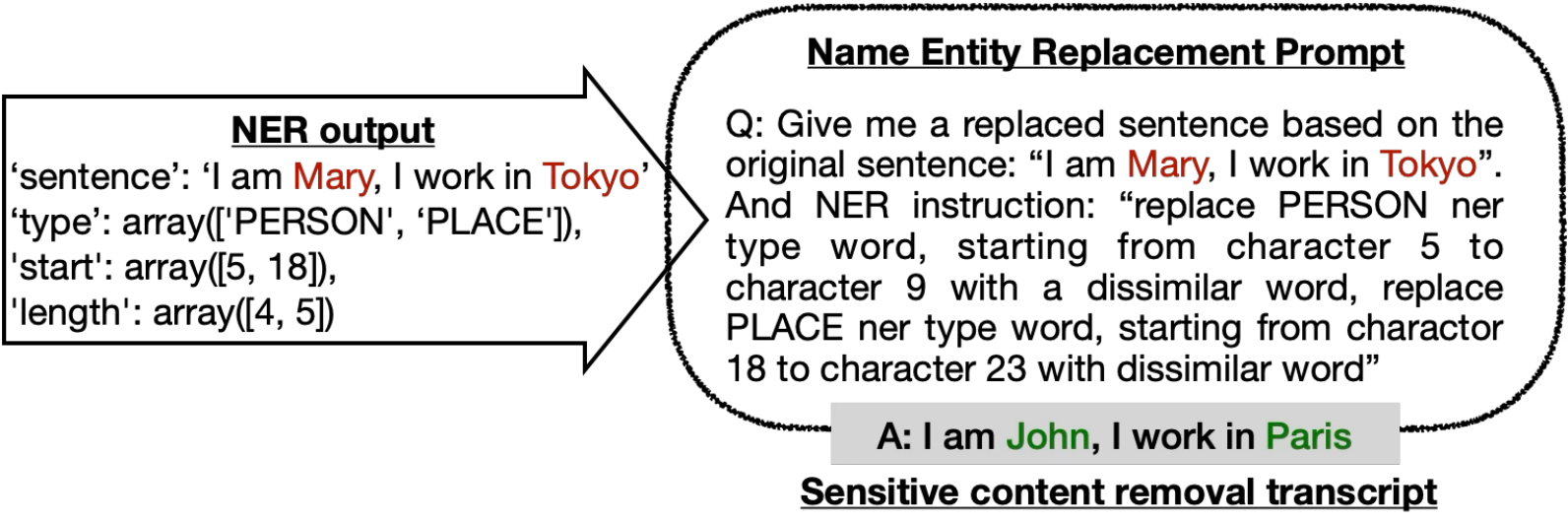}
    \caption{Illustration of User Prompt Used in Proposed Pipeline.}
    \label{fig:Illustration of Prompts}

\end{figure}

To increase the output accuracy of the LLM, several principles of prompt engineering were used when designing the system prompt, such as simplifying complex tasks, few-shot prompting, usage of delimiters\cite{bsharat2023principled}.
Figure \ref{fig:Illustration of Prompts} shows the specifics of the prompt. The NER output is converted into a human-readable, LLM-friendly sentence derived from the original NER model output.

\subsection{Speaker Identity Concealment}
\label{sec:spk-description-gen}
Most DNN-based speaker anonymization systems rely on pre-trained speaker encoders \cite{desplanques2020ecapa} or Neural Audio Codec Language Models \cite{10447871} to extract speaker-related representations. These systems either transform the source speaker embedding \cite{meyer2022anonymizing,miao2023language,MUSA} or replace it with embeddings from an external pool \cite{fang2019speaker, miao22_odyssey, championdeep, 10447871,miao2025benchmark} to remove source speaker identity clues. Both methods involve combining anonymized speaker and content embeddings, which are then used by a speech generator to produce anonymized speech. Recent advancements in prompt-based text-to-speech models allow for control over speaker attributes via text, reducing the reliance on vocal characteristics and minimizing re-identification risks, while enabling greater diversity in synthetic speech.

How to produce speaker descriptions depends on the specific prompt-based text-to-speech model selected. In this paper, we utilize Parler-TTS Large v1\footnote{\url{https://huggingface.co/parler-tts/parler-tts-large-v1}}.
This TTS system employs a decoder-only Transformer architecture. 
The input consists of: 1) Transcript text: The text to be synthesized, pre-pended to the input sequence. 2) Speaker description text: natural language descriptions of desired speech characteristics. 
This is processed by a pre-trained T5 \cite{T5} text encoder and fed to the decoder through cross-attention \cite{chen2021crossvit}.
The output of the decoder is a sequence of discrete tokens representing audio features, which are then converted to audio waveforms using the Descript Audio Codec decoder \cite{kumar2023highfidelity}.
The model is trained on approximately 45K hours of audio data. During training, the model is optimized to generate speech that accurately reflects both the provided text transcript and speaker description.

As for speaker prompt of Parler-TTS, it is a speaker description 
specifying various attributes, such as gender, pitch, speaking rate, and recording conditions that can be adjusted to provide varying audio sets\footnote{e.g., specific speaker - '\textbf{Jon's} reads a book with a quick and monotone voice. The recording is close-sounding and clean.' or random speaker - '\textbf{a male} reads a book slowly with very high-pitched and expressive and animated voice. The recording is distant-sounding.'}. 
The attributes will be described and investigated in Section \ref{sec:results}.

\subsection{Speech Generation - Parler-TTS} 
After obtaining the speaker description and the sensitive content removal transcript prompt, these are fed into Parler-TTS to generate anonymized speech. 

\section{Evaluation}
\label{sec:eva}

\subsection{Evaluation Dataset and Preprocessing}
The experiment uses a benchmark dataset called SLUE-VoxPopuli \cite{shon2022slue}, which can be used for ASR and NER tasks. The named entities in SLUE-VoxPopuli include PLACE, QUANT, ORG, WHEN, NORP, PERSON, and LAW.
The evaluation dataset is selected from a total of 456 speakers with 6,842 utterances in the training and testing sets. After removing utterances with duplicate transcripts and speakers with fewer than 10 utterances\footnote{More optimal for objective evaluation}, 161 speakers with 3,729 utterances are obtained.

\subsection{Evaluation Metrics}
In terms of anonymized speech evaluation, we use two privacy metrics to assess content and speaker privacy protection and two utility metrics to assess speech intelligibility and quality of anonymized speech.
\subsubsection{Privacy Evaluation}
\paragraph{Content Privacy Metrics}
\label{sec:content-eva}
Content privacy ability depends on the performance of the cascade models of ASR, NER, and named entity replacement. We use word error rate (WER) of the original speech, F1 score of NER on the predicted transcription, and named entity replacement accuracy, respectively, as shown in the dotted box on the right side of the Figure \ref{fig:Illustration of Pipeline}. WER is measured using an ASR evaluation model denoted as $ASR_\text{eval}$, which is the same model used for speech transcription introduced in Section \ref{sec:asr}. Lower WER, higher F1, and higher replacement accuracy indicate better content protection.

\paragraph{Speaker Privacy Metric - False Acceptance Rate}
To evaluate the effectiveness of speaker privacy protection in anonymized speech, we simulate an attacker who uses an automatic speaker verification (ASV) model ($ASV_\text{eval}$) to infer the source speaker's identity from the anonymized speech. 
Conceptually, the attacker compares the anonymized speech (i.e., ASV probe) with an unprotected utterance from the targeted speaker (i.e., ASV enrollment)\footnote{We assume that the attack has access to unprotected utterances, for example, retrieved from social network services.}. If the probe and enrollment are sufficiently similar, the attacker decides that the anonymized speech is uttered by the target speaker. 

In implementation, the probe and enrollment utterances are fed to the attacker's ASV model. Speaker representations are extracted, and their cosine similarity is compared with a threshold for decision making, i.e, whether the probe utterance matches the target speaker.
Following the convention in Voice Privacy challenge~\cite{vpc2022sum}, we treat the case as \emph{false acceptance} when the attacker `correctly' decides that the anonymized probe matches the target speaker\footnote{In Voice Privacy challenge~\cite{vpc2022sum}, given an reference from a target speaker, the anonymized probe and the unprotected probe are treated as being from the \emph{negative} and \emph{positive} classes, respectively. A false acceptance is made when the anonymized speech (from negative class) is `incorrectly' classified as being positive, even though the decision is correct from the attacker's perspective.}.  

The decision threshold of the attacker is determined using original speech pairs: positive pairs (from the same speaker) and negative pairs (from different speakers). 
Its value is set at the level where the false acceptance and false rejection rates are equal\footnote{It corresponds to the decision threshold achieving the equal error rate \cite{vpc2022sum}.}. 
The false acceptance rate (FAR) of the anonymized speech is then calculated as the ratio of the number of falsely accepted anonymized utterances to the total number of anonymized utterances, as shown in the dotted box at top middle of the Figure \ref{fig:Illustration of Pipeline}.
A lower FAR indicates effective anonymization, i.e., the attacker has a lower success rate of identifying the target speakers.
The attacker's $ASV_\text{eval}$ model is the ECAPA-TDNN \cite{desplanques2020ecapa}, publicly available\footnote{\url{https://huggingface.co/speechbrain/spkrec-ecapa-voxceleb}}.

\subsubsection{Utility Evaluation}

\paragraph{Speech Intelligibility Utility Metric - Word Error Rate}
To evaluate the preservation of \textit{speech content} in anonymized speech, we compute WER using $ASR_\text{eval}$ as shown in the dotted box at top middle of the Figure \ref{fig:Illustration of Pipeline}. A lower WER of anonymized speech, comparable to that of the original speech, suggests effective preservation of the speech content.

\paragraph{Speech Quality Utility Metric - Predicted Mean Opinion Score}
 To quantify the quality of the anonymized speech, we calculate objective Predicted Mean Opinion Score (PMOS) using Torchaudio-Squim model \cite{kumar2023torchaudio}, as shown in the dotted box at the top middle of the Figure \ref{fig:Illustration of Pipeline}. A higher PMOS, similar to that of the original speech, indicates a good speech quality preservation ability\footnote{We did not do subjective listening tests because the cost is beyond budget.}.

\subsection{Evaluation Baseline}

We compare the proposed prompt-based model with VALL-E\cite{wang2023neural} \footnote{https://github.com/Plachtaa/VALL-E-X} and XTTS-v2 \cite{XTTS}\footnote{https://huggingface.co/coqui/XTTS-v2}, both of which are popular zero-shot TTS models that generate speech by conditioning on text and speaker audio to mimic the speaker's identity and speaking style. The key difference is that the zero-shot TTS models rely on actual speaker's speech, whereas the proposed prompt-based system uses text-based edits of speaker attributes to control the speaker's identity in the generated audio.

\begin{table}[t]
 \centering
  \footnotesize
  \setlength{\tabcolsep}{3pt}
 \begin{tabular}{l|ccc}
 \toprule
 & FAR & WER  & PMOS   \\
 \midrule
Original & 100.0  & 23.67 &  4.48 \\
VALL-E  \cite{wang2023neural} & 0.0 & 33.69 & 4.01 \\
XTTS-v2 \cite{XTTS}   &  0.0 & 14.17  & 3.99 \\
Prompt-based   & 0.0 & 15.50 &  4.01 \\
    \bottomrule
 \end{tabular}
  \vspace{4mm}
  \caption{FAR (\%) $\downarrow$, WER (\%) $\downarrow$, PMOS $\uparrow$ for different systems. Original and anonymized speech share the same text transcription.}
\label{tab:Evaluation-General}
  \vspace{-2mm}
\end{table}

\subsection{Experimental Results}
\label{sec:results}

\subsubsection{The Results of the Proposed and Baseline Systems}

First of all, regardless of the TTS model used, ASR and NER components as described in Section \ref{sec:asr}, Sections \ref{sec:ner}, and Sections \ref{sec:ner-replace} are essential for removing sensitive content, which is then used as the text input for the TTS models. The wav2vec 2.0 + CTC model achieves a WER of 19.00\% for original speech transcription, while the DeBERTa-L model achieves an F1 score of 71.80\% for NER on predicted transcriptions, and named entity replacement reaches an accuracy of 99.95\%. These results demonstrate acceptable performance in content transcription, named entity recognition, and robust replacement capabilities, thereby ensuring satisfactory content protection. In the following experiments, 
when calculating the WER of anonymized speech, we use the transcripts with sensitive content removed as the ground truth.

Then, we examine the utility and privacy of the anonymized speech generated by different TTS models, either using speaker speech (VALL-E, XTTS-v2) or speaker descriptions (prompt-based).
The results are listed in Table \ref{tab:Evaluation-General}. 
The FAR of the original speech is 100\%, as the original speech lacks any protection. In contrast, the FAR of anonymized speech for VALL-E, XTTS-v2, and prompt-based system is 0\%, indicating strong privacy protection. This is not surprising because the TTS pipeline trained on data of speakers who are disjoint from the target speakers in the evaluation set. The anonymized voice produced by TTS models is expected to be different from the target speakers. This is different from systems that produce an anonymized voice given the speaker representation of the input utterance to be anonymized. When the anonymized voice depends on the input, there is a chance of re-identify the speaker from the anonymized utterance~\cite{championdeep}.

In terms of WER,  except for the anonymized speech generated by VALLE-X, the original speech has higher WERs than anonymized speech, a trend also observed in other TTS studies \cite{alharthi24_syndata4genai}. This difference may be attributed to the direct generation of TTS audio from text, which produces high-quality audio and achieves lower WER. Additionally, named entity replacement often replaces simpler words, which are easier to generate accurately. 

For PMOS, the original speech achieves a better or similar score compared to anonymized speech. 
Overall, the anonymized speech generated by XTTS-v2 and the prompt-based method demonstrate robust speaker privacy protection along with good speech intelligibility and quality. However, the prompt-based approach, which uses text-based edits of speaker attributes, is more flexible, transparent, and user-friendly compared to methods that rely on hard-coded speaker characteristics.

\begin{table}[h]
 \centering
  \footnotesize
 \begin{tabular}{l|lcc}
 \toprule
 Subcategories &  WER  & PMOS  \\
 \midrule
   American   & \cellcolor[rgb]{0.91, 0.91, 0.91} 16.00 & \cellcolor[rgb]{0.83, 0.83, 0.83} 4.17 \\ 
   Brazilian   & \cellcolor[rgb]{0.86, 0.86, 0.86} 17.68 & \cellcolor[rgb]{0.73, 0.73, 0.73} 4.29 \\
   Bulgarian   & \cellcolor[rgb]{0.93, 0.93, 0.93} 15.31 & \cellcolor[rgb]{0.70, 0.70, 0.70} 4.32 \\
   Catalan    & \cellcolor[rgb]{0.91, 0.91, 0.91} 16.03 & \cellcolor[rgb]{1.00, 1.00, 1.00} 3.86 \\
   Croatian   & \cellcolor[rgb]{0.91, 0.91, 0.91} 16.00 & \cellcolor[rgb]{0.83, 0.83, 0.83} 4.17 \\
   Dutch     & \cellcolor[rgb]{0.94, 0.94, 0.94} 14.89 & \cellcolor[rgb]{0.95, 0.95, 0.95} 3.98 \\
    Estonian   & \cellcolor[rgb]{0.88, 0.88, 0.88} 17.16 & \cellcolor[rgb]{0.86, 0.86, 0.86} 4.12 \\
    French    & \cellcolor[rgb]{0.90, 0.90, 0.90} 16.33 & \cellcolor[rgb]{0.81, 0.81, 0.81} 4.19 \\ 
   Hungarian   & \cellcolor[rgb]{0.95, 0.95, 0.95} 14.73 & \cellcolor[rgb]{0.59, 0.59, 0.59} 4.42 \\ 
   Indonesian  & \cellcolor[rgb]{0.92, 0.92, 0.92} 15.76 & \cellcolor[rgb]{0.94, 0.94, 0.94} 4.00 \\
      Italian    & \cellcolor[rgb]{0.59, 0.59, 0.59} 23.76 & \cellcolor[rgb]{0.86, 0.86, 0.86} 4.13 \\
   Japanese   & \cellcolor[rgb]{0.94, 0.94, 0.94} 14.87 & \cellcolor[rgb]{0.82, 0.82, 0.82} 4.18 \\
  Lithuanian   & \cellcolor[rgb]{0.93, 0.93, 0.93} 15.40 & \cellcolor[rgb]{0.81, 0.81, 0.81} 4.19 \\
    North Irish   & \cellcolor[rgb]{0.88, 0.88, 0.88} 16.96 & \cellcolor[rgb]{0.85, 0.85, 0.85} 4.14 \\
  Polish    & \cellcolor[rgb]{0.95, 0.95, 0.95} 14.63 & \cellcolor[rgb]{0.88, 0.88, 0.88} 4.10 \\ 
   Scottish   & \cellcolor[rgb]{0.95, 0.95, 0.95} 14.59 & \cellcolor[rgb]{0.68, 0.68, 0.68} 4.34 \\
   Slovene    & \cellcolor[rgb]{0.93, 0.93, 0.93} 15.37 & \cellcolor[rgb]{0.68, 0.68, 0.68} 4.34 \\ 
   South England  & \cellcolor[rgb]{0.91, 0.91, 0.91} 16.15 & \cellcolor[rgb]{0.78, 0.78, 0.78} 4.23\\
  Vietnamese  & \cellcolor[rgb]{0.88, 0.88, 0.88} 17.15 & \cellcolor[rgb]{0.68, 0.68, 0.68} 4.34 \\
    
    Australian  & \cellcolor[rgb]{0.98, 0.98, 0.98} 13.04 & \cellcolor[rgb]{0.95, 0.95, 0.95} 3.99  \\
    British    & \cellcolor[rgb]{0.96, 0.96, 0.96} 13.95 & \cellcolor[rgb]{0.81, 0.81, 0.81} 4.20\\ 
    Canadian   & \cellcolor[rgb]{0.93, 0.93, 0.93} 15.52 & \cellcolor[rgb]{0.79, 0.79, 0.79} 4.22\\ 
    Chinese    & \cellcolor[rgb]{0.97, 0.97, 0.97} 13.66 & \cellcolor[rgb]{0.80, 0.80, 0.80} 4.21\\ 
    Czech     & \cellcolor[rgb]{0.95, 0.95, 0.95} 14.42 & \cellcolor[rgb]{0.88, 0.88, 0.88} 4.09\\ 
    Egyptian   & \cellcolor[rgb]{0.85, 0.85, 0.85} 17.93 & \cellcolor[rgb]{0.97, 0.97, 0.97} 3.93\\ 
    Finnish    & \cellcolor[rgb]{0.90, 0.90, 0.90} 16.22 & \cellcolor[rgb]{0.86, 0.86, 0.86} 4.12\\ 
    German    & \cellcolor[rgb]{0.97, 0.97, 0.97} 13.56 & \cellcolor[rgb]{0.80, 0.80, 0.80} 4.21\\
    Indian    & \cellcolor[rgb]{0.95, 0.95, 0.95} 14.79 & \cellcolor[rgb]{0.86, 0.86, 0.86} 4.12\\ 
    Irish     & \cellcolor[rgb]{0.96, 0.96, 0.96} 13.88 & \cellcolor[rgb]{0.81, 0.81, 0.81} 4.19\\ 
    Jamaican   & \cellcolor[rgb]{0.95, 0.95, 0.95} 14.65 & \cellcolor[rgb]{0.73, 0.73, 0.73} 4.29\\ 
    Latin American & \cellcolor[rgb]{0.94, 0.94, 0.94} 15.25 & \cellcolor[rgb]{0.86, 0.86, 0.86} 4.13\\ 
    North England  & \cellcolor[rgb]{0.97, 0.97, 0.97} 13.62 & \cellcolor[rgb]{0.86, 0.86, 0.86} 4.12\\ 
    Pakistani   & \cellcolor[rgb]{0.97, 0.97, 0.97} 13.37 & \cellcolor[rgb]{0.75, 0.75, 0.75} 4.27\\ 
    Romanian   & \cellcolor[rgb]{0.91, 0.91, 0.91} 16.14 & \cellcolor[rgb]{0.96, 0.96, 0.96} 3.96\\
    Slovak    & \cellcolor[rgb]{1.00, 1.00, 1.00} 12.07 & \cellcolor[rgb]{0.85, 0.85, 0.85} 4.14\\ 
    South African & \cellcolor[rgb]{0.98, 0.98, 0.98} 13.20 & \cellcolor[rgb]{0.92, 0.92, 0.92} 4.09\\
    Spanish    & \cellcolor[rgb]{0.93, 0.93, 0.93} 15.34 & \cellcolor[rgb]{0.94, 0.94, 0.94} 4.00  \\
    Wales     & \cellcolor[rgb]{0.86, 0.86, 0.86} 17.48 & \cellcolor[rgb]{0.77, 0.77, 0.77} 4.24 \\

 \bottomrule
\end{tabular}
 \vspace{4mm}
 \caption{WER (\%) $\downarrow$, PMOS $\uparrow$ for different accent description variants. A darker cell color indicates a higher metric value in each column. FAR for original speech is 100\%, indicating no privacy protection. After anonymization, FAR changes to 0\%, indicating strong privacy protection. For simplicity, we omit FAR in the table.}
\label{tab:accent}
  \vspace{-2mm}
\end{table}

\begin{figure}[t]
    \centering
   \includegraphics[width=1\linewidth, trim=10 15 10 0]{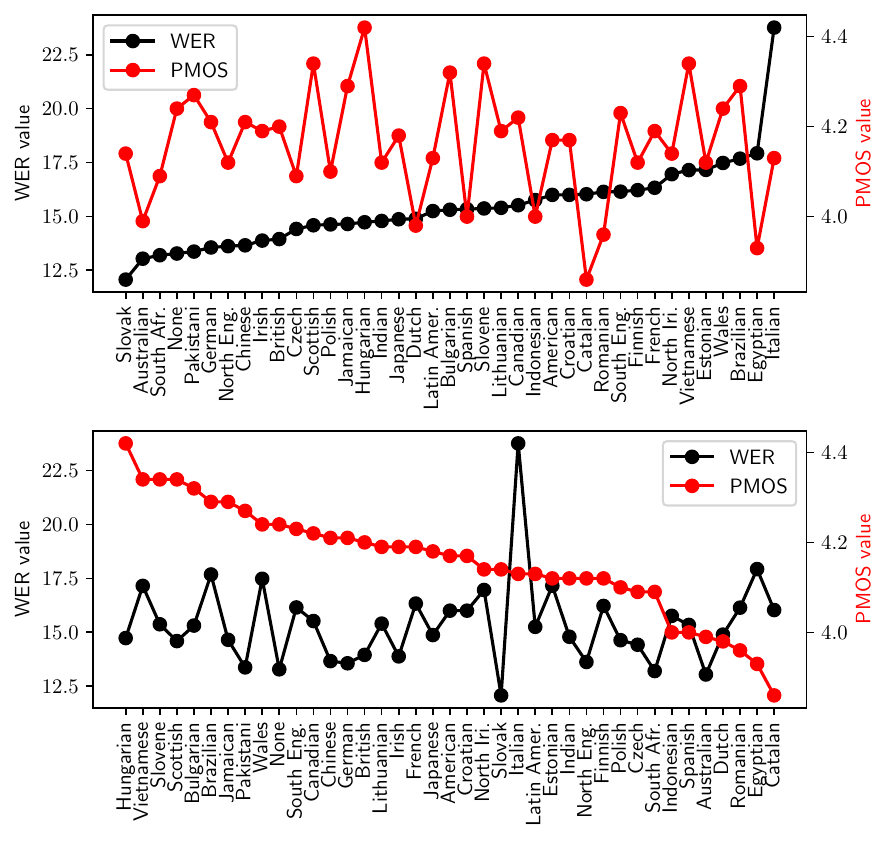}
            \vspace{1mm}
    \caption{Sorted WER (top) and sorted PMOS (bottom) for different accent description variants.}
    \label{fig:accent-plot}
\end{figure}

\begin{table}[h]
 \centering
 \footnotesize
 \begin{tabular}{c|l|cc}
 \toprule
 Attributes & Subcategories & WER & PMOS \\
 \midrule
 \multirow{2}{*}{\rotatebox{0}{\scriptsize Gender}}&      Female       & \cellcolor[rgb]{0.89, 0.89, 0.89} 14.85 & \cellcolor[rgb]{0.85, 0.85, 0.85} 4.10\\ 
  &       Male        & \cellcolor[rgb]{0.75, 0.75, 0.75} 16.88 & \cellcolor[rgb]{0.88, 0.88, 0.88} 4.07\\ 

 \midrule
 \multirow{5}{*}{\rotatebox{0}{Pitch}} & Very low-pitched  & \cellcolor[rgb]{0.91, 0.91, 0.91} 14.51 & \cellcolor[rgb]{0.92, 0.92, 0.92} 4.03\\ 
  &    Low-pitched    & \cellcolor[rgb]{0.97, 0.97, 0.97} 13.25 & \cellcolor[rgb]{0.82, 0.82, 0.82} 4.12\\ 
  &      Normal       & \cellcolor[rgb]{0.93, 0.93, 0.93} 14.10 & \cellcolor[rgb]{0.64, 0.64, 0.64} 4.25\\ 
  &   High-pitched    & \cellcolor[rgb]{0.81, 0.81, 0.81} 16.01 & \cellcolor[rgb]{0.90, 0.90, 0.90} 4.05\\ 
  & Very high-pitched & \cellcolor[rgb]{0.84, 0.84, 0.84} 15.62 & \cellcolor[rgb]{0.65, 0.65, 0.65} 4.24\\ 
 
 \midrule
 \multirow{3}{*}{\rotatebox{0}{\shortstack{Pitch\\modula.}}}

  &  Express. anim.   & \cellcolor[rgb]{0.88, 0.88, 0.88} 15.06 & \cellcolor[rgb]{0.67, 0.67, 0.67} 4.23\\ 
  &     Monotone      & \cellcolor[rgb]{0.86, 0.86, 0.86} 15.26 & \cellcolor[rgb]{0.72, 0.72, 0.72} 4.20\\ 
  &       Empty       & \cellcolor[rgb]{0.92, 0.92, 0.92} 14.38 & \cellcolor[rgb]{0.85, 0.85, 0.85} 4.10\\ 
 \midrule
 \multirow{5}{*}{\rotatebox{0}{\shortstack{Channel\\condition}}}  &       Clean       & \cellcolor[rgb]{0.90, 0.90, 0.90} 14.60 & \cellcolor[rgb]{0.83, 0.83, 0.83} 4.11\\ 
  &    Close-sound    & \cellcolor[rgb]{0.90, 0.90, 0.90} 14.60 & \cellcolor[rgb]{0.90, 0.90, 0.90} 4.05\\ 
  &   Distant-sound   & \cellcolor[rgb]{0.81, 0.81, 0.81} 16.10 & \cellcolor[rgb]{0.61, 0.61, 0.61} 4.27\\ 
  &       Noisy       & \cellcolor[rgb]{0.83, 0.83, 0.83} 15.83 & \cellcolor[rgb]{0.74, 0.74, 0.74} 4.19\\ 
  &      Normal       & \cellcolor[rgb]{0.90, 0.90, 0.90} 14.74 & \cellcolor[rgb]{0.59, 0.59, 0.59} 4.28\\ 

 \midrule
 
  \multirow{5}{*}{\rotatebox{0}{\shortstack{Speaking\\rate}}}
  &    Very slowly    & \cellcolor[rgb]{0.84, 0.84, 0.84} 15.64 & \cellcolor[rgb]{0.87, 0.87, 0.87} 4.08\\ 
  &      Slowly       & \cellcolor[rgb]{0.86, 0.86, 0.86} 15.25 & \cellcolor[rgb]{0.59, 0.59, 0.59} 4.28\\ 
  &     Normally      & \cellcolor[rgb]{1.00, 1.00, 1.00} 12.39 & \cellcolor[rgb]{0.67, 0.67, 0.67} 4.23\\ 
  &      Quickly      & \cellcolor[rgb]{0.85, 0.85, 0.85} 15.55 & \cellcolor[rgb]{1.00, 1.00, 1.00} 3.91\\ 
  &   Very quickly    & \cellcolor[rgb]{0.59, 0.59, 0.59} 18.56 & \cellcolor[rgb]{0.65, 0.65, 0.65} 4.24\\ 
 
 \bottomrule
 \end{tabular}
 \vspace{4mm}
 \caption{WER (\%) $\downarrow$, PMOS $\uparrow$ for different gender, pitch, pitch modulation, channel condition and speaking rate description variants. A darker cell color indicates a higher metric value in each column. FAR for original speech is 100\%, indicating no privacy protection. After anonymization, FAR changes to 0\%, indicating strong privacy protection. For simplicity, we omit FAR in the table.}
 \label{tab:gender}
   \vspace{-2mm}
\end{table}

\subsubsection{Investigations on Altering Speech Characteristics in Speaker Descriptions for Prompt-based System}

To further evaluate the proposed prompt-based pipeline, the speech generated uses speaker descriptions with randomly generated configurations. The numerous combinations also lead to questions on how each unique attribute can potentially affect the eventual utility and privacy of the anonymized audio. As such, an experiment was conducted by generating audio sets while only varying one attribute and randomly setting the other attributes.
There are six attributes: accent, gender, channel condition, pitch modulation, speaking rate, and pitch. Each attribute has various subcategories. For example, the gender attribute has two subcategories: female and male. For each subcategory, such as male, 40 random combinations of attributes—'Male - random\{accent, channel condition, pitch modulation, speaking rate, pitch\}'—are generated, with 10 utterances per combination, resulting in a total of 400 speaker descriptions for each subcategory.

The evaluation results for FAR, WER, and PMOS for each attribute and its subcategories are shown in Tables \ref{tab:accent} and \ref{tab:gender}, and Figure \ref{fig:accent-plot}. 
For FAR, all variations produce a FAR of 0\%. This is consistent with the pipeline generation results. Within the subcategories, there are differences in the WER and PMOS which can affect the utility of the audio. For example, it would be better to avoid the Italian accent compared to the other accents due to its relatively higher WER (23.7\%.) despite having good audio quality as seen from its PMOS (4.13). As the main objective is to anonymize speaker's identity and sensitive content, it is still important to ensure the retention of all other content. For the remaining subcategories that have WER ranging between 12\% to 19\%, it may be more beneficial to select subcategories on the lower WER end, such as choosing the Slovak or Australian accent, female, low-pitched, clean, close-sounding, normally spoken rate speech. For speech-related training tasks that require good audio quality, it would be preferable to avoid the subcategories below PMOS of 4.00 such as Catalan, Egyptian, Romanian, Dutch accents, and very quickly speaking speech.

\section{Discussion}
\label{sec:dis}

\noindent
\textbf{Practical Content Anonymization:}
This paper follows the common settings of ASR + NER + LLM to detect and replace sensitive content. However, there are several concerns:
(i) This approach still suffers from unavoidable ASR encoding errors (19.00\% WER on original speech) and NER detection errors (71.80\% F1 score on predicted transcriptions), which affect the accuracy of content protection.
(ii) The NER approach assumes that "sensitive content" is limited to named entities. This is a narrow definition, though it aligns with common settings in the known challenges of speech content protection. The process could be streamlined by using a single LLM with carefully designed content prompts, enabling more flexible and controlled replacements.
(iii) Current content privacy evaluations rely on WER, which, while useful for measuring word-level substitutions, does not account for context, long-term relationships, or reasoning that may still reveal the speaker's identity. Hence, the conventional WER metric is insufficient to evaluate content privacy protection, as it doesn't capture the semantic quality of substitutions.

\noindent
\textbf{Irreversible Anonymization:}
The current pipeline is designed for irreversible anonymization to ensure strong privacy guarantees.
However, reversibility is required for certain use cases. For example, in forensic or security contexts, the original content and speaker identity information should be recoverable to facilitate the investigation. A practical solution could involve allowing users or authorized parties to maintain a secure mapping between the original and anonymized speaker/content, enabling recovery when necessary.

\noindent
\textbf{Prosody Preservation:}
The current pipeline cannot preserve prosody while ensuring strong speaker protection, as prosody can reveal speaker identity to some extent. However, it is promising to add a prosody preservation branch on top of the current design, e.g., by disentangling prosody information from the original speech and mimicking it.

\noindent
\textbf{Controllability of the Anonymized Speaker Identity:}
We have conducted experiments and found that due to the generative nature of prompt-based methods, even when using the same description, the output may sound like different persons across generations. To ensure consistent identity control, Parler-TTS also supports explicit speaker specification, which can be used directly for deterministic voice conversion. 

\section{Conclusion}
The paper presents a prompt-based speaker and content anonymization pipeline by leveraging the SOTA ASR, NER, Llama-3.2 and Parler-TTS model. The pipeline effectively mitigates the two concerns mentioned at the start of the paper: 1) The potential of identity theft: The use of Parler-TTS enables the generation of diverse pseudo-speakers controlled by speaker descriptions, without relying on any original speaker characteristics, ensuring that the original voiceprint can no longer be stolen. 2) The potential of re-identification of speaker from content: The ASR transcription and name entity recognition and replacement achieves acceptable performance, ensuring a significant number of sensitive content has been removed. An investigation into variations of speaker descriptions also demonstrated how each subcategory can affect the utility of the audio for future task applications. 

This is the first work to use prompt-based edits of speaker attributes for dual anonymization. Despite the progress, many interesting aspects remain to be investigated as section \ref{sec:dis} disscussed, and we expect that research in this direction will continue to evolve and attract greater attention in the future
{\small
\bibliographystyle{ieee}
\bibliography{ref}
}

\end{document}